\documentclass[aps,superscriptaddress,floatfix,twocolumn,footinbib]{revtex4-1}
\usepackage{amssymb}
\usepackage{amsmath}
\usepackage{amsfonts}
\usepackage{appendix}
\usepackage{bm}
\usepackage{graphicx}
\usepackage{epsfig}
\usepackage{epstopdf}
\usepackage{balance}
\usepackage[dvipsnames]{xcolor}
\usepackage{calc}
\usepackage{natbib}
\usepackage[colorlinks,
            linkcolor=blue,
            anchorcolor=blue,
            citecolor=blue,
            urlcolor=blue]{hyperref}
\usepackage{lipsum}
\usepackage[version=3]{mhchem}
\usepackage{diagbox}

\usepackage{color}
\usepackage{soul}

\usepackage{setspace}
\usepackage{bbding}
\usepackage[colorlinks,linkcolor=blue]{hyperref}

\begin{document}
\title{Efficient Method for Prediction of Meta-stable/Ground Multipolar Ordered States and its Application in Monolayer $\alpha$-\ce{RuX3} (X=Cl,I)}

\author{Wen-Xuan Qiu}
\affiliation{School of Physics and Wuhan National High Magnetic field center,
Huazhong University of Science and Technology, Wuhan 430074,  China}
\author{Jin-Yu Zou}
\affiliation{School of Physics and Wuhan National High Magnetic field center,
Huazhong University of Science and Technology, Wuhan 430074,  China}
\author{Ai-Yun Luo}
\affiliation{School of Physics and Wuhan National High Magnetic field center,
Huazhong University of Science and Technology, Wuhan 430074,  China}
\author{Zhi-Hai Cui}
\affiliation{Institute of Physics, Chinese Academy of Sciences, Beijing 100190, China.}
\author{Zhi-Da Song}
\affiliation{Institute of Physics, Chinese Academy of Sciences, Beijing 100190, China.}
\affiliation{Department of Physics, Princeton University, Princeton, New Jersey 08544, USA}
\author{Jin-Hua Gao}
\affiliation{School of Physics and Wuhan National High Magnetic field center,
Huazhong University of Science and Technology, Wuhan 430074,  China}
\author{Yi-Lin Wang}\email{yilinwang@ustc.edu.cn}
\affiliation{Hefei National Laboratory for Physical Sciences at Microscale, University of Science and Technology of China, Hefei, Anhui 230026, China}
\author{Gang Xu}
\email{gangxu@hust.edu.cn}
\affiliation{School of Physics and Wuhan National High Magnetic field center,
Huazhong University of Science and Technology, Wuhan 430074,  China}

\begin{abstract}
Exotic high-rank multipolar order parameters have been found to be unexpectedly active in more and more correlated materials in recent years. Such multipoles are usually dubbed as ``Hidden Orders'' since they are insensitive to common experimental probes. Theoretically, it is also difficult to predict multipolar orders via \textit{ab initio} calculations in real materials. Here, we present an efficient method to predict possible multipoles in materials based on linear response theory under random phase approximation. Using this method, we successfully predict two pure meta-stable magnetic octupolar states in monolayer $\alpha$-\ce{RuCl3}, which is confirmed by self-consistent unrestricted Hartree-Fock calculations. We then demonstrate that these octupolar states can be stabilized in monolayer $\alpha$-\ce{RuI3}, one of which becomes the octupolar ground state. Furthermore, we also predict a fingerprint of orthogonal magnetization pattern produced by the octupole moment, which can be easily detected by experiment. The method and the example presented in this work serve as a guidance for searching multipolar order parameters in other correlated materials.
\end{abstract}

\maketitle
\section{introduction}
High-rank multipolar order parameters (OPs), induced by multiple orbital degrees of freedom themselves or their coupling to spin sector via strong spin-orbital coupling (SOC), have been found to be unexpectedly active in more and more correlated materials such as $4f$, $5f$ and $5d$ systems in recent years~\cite{tokunaga2006nmr,tokunaga2006multipolar,hiroaki2008,revaps2009,revjpsj2009,chen2010,chen2011,ikeda2012,witczak2014,harter2017,suzuki2017cluster,suzuki2018,hayami2018microscopic,hayami2018classification,watanabe2018,ishikawa2019,ggaulin2020,gaulin2020,hirai2020}, although they are usually considered as weak terms comparing to dipoles under multipole expansion. In some cases, they can even act as the primary OPs. One of such examples is the famous ``Hidden Order'' (HO) phase transition occurring around 17.5 K in URu$_2$Si$_2$~\cite{palstra1985,ikeda1998theory,elgazzar2009hidden,oppeneer2010electronic,mydosh2011,oppeneer2011spin,okazaki2011rotational,tonegawa2012cyclotron,meng2013imaging,mydosh2014}, and many different kinds of multipolar moments, such as quadrupole~\cite{santini1994,santini1998,ohkawa1999}, octupole~\cite{kiss2005,hanzawa2007}, hexadecapole~\cite{haule2009,kusunose2011,kung2015,kung2016} and dotriacontapole~\cite{cricchio2009,ikeda2012,ikeda2014multipole} have been suggested to be the primary OPs in this HO phase. Different from the conventional dipoles, much richer and exotic orders and low-energy excitations could be expected arising from multipolar OPs due to their higher degrees of freedom. For instance, superconductivity mediated by multipole fluctuation~\cite{ikeda1998theory,kotegawa2003,koga2006,goto2011,matsubayashi2012,ikeda2014multipole,kittaka2014multiband,ikeda2015emergent,nomoto2016,sumita2016,kattori2017local,yamashita2017fully,bai2021}, multipolar Kondo effects~\cite{cox1987,cox1988,yatskar1996,cox1998,haule2009,onimaru2016,yamane2018,patri2020} with exotic non-Fermi-liquid fixed points~\cite{patri2020}, cross-correlated responses~\cite{popov1999,hur2004,lorenz2004,rai2007,chikara2009,hayami2014,hayami2014spontaneous,satoru2015,hayami2015,satoru2015quantum,khanh2016,hayami2016emergent,hayami2016asymmetric,matsumoto2017,suzuki2017cluster,yanagi2017optical,ikhlas2017large,hayami2018,hayami2018microscopic,hayami2018mean,yanagi2018theory,yanagi2018manipulating,florian2018,shitade2018} have been found. Therefore, the important roles played by multipolar OPs are attracting extensive attentions and are considered as significant factors to interpret some exotic physical phenomena~\cite{revaps2009,revjpsj2009,witczak2014,watanabe2018}.

However, such high-rank OPs pose a big challenge to experimental detections, since they are not or weakly coupled to the common experimental probes~\cite{wang2017}, or they are usually accompanied with a primary dipolar OP~\cite{amitsuka2010,walker2011,dos2016,tian2017,wang2017} that dominates the experimental signals. This is the reason why the multipolar OPs are usually dubbed as HOs and their roles are rarely recognized even though they might be widely present in materials. Theoretically, predicting multipolar OPs in real materials from \textit{ab initio} calculations~\cite{kresse1996,ghosh2005electronic,shick2005effect,haule2009,cricchio2009,suzuki2009dynamical,elgazzar2009hidden,suzuki2010change,suzuki2010,derlind2010computational,oppeneer2010electronic,modin2011indication,ikeda2012,suzuki2013,suzuki2014,werwi2014exceptional,goho2015,maldonado2016crystal,suzuki2018,huebsch2020} is crucial but also not an easy task, since (1) most of these materials involve both strong SOC and electronic correlations that should be properly treated by methods such as the self-consistent unrestricted Hartree-Fock mean-field (HFMF) method with all the off-diagonal terms of local density matrix kept~\cite{yilin2017}; (2) self-consistent calculations of the multipolar states extremely depend on the transcendental knowledge of the possible multipoles and the related symmetry breaking of Hamiltonian to induce the desired OPs; (3) the energy differences between different multipoles are usually very tiny. As a result, such calculations should be performed many times with very high numerical accuracy such that they take too much time and even become unfeasible in systems with a lot of atoms. Thus, a highly efficient method to search for all the possible multipolar states is very desirable, and the predictions of their fingerprints in physical observables that can be easily measured experimentally are also very important to uncover HO physics in materials.

In this work, based on linear response theory (LRT) under random phase approximation (RPA)~\cite{hiroaki2008,ikeda2012,jishi2013,ikeda2014multipole,ikeda2015emergent,kattori2017local,suzuki2018}, we develop a numerical method starting from the density functional theory (DFT) calculations, to search for all possible multipolar OPs efficiently for the spin-orbital entangled correlated electronic materials, which only requires a fast single-shot calculation. We use monolayer $\alpha$-\ce{RuX3} (X=Cl,I) as an example to demonstrate its formalism, capabilities and effectiveness. It has correctly reproduced the Zigzag magnetic ground state of $\alpha$-\ce{RuCl3} as found in the neutron scattering experiments~\cite{sears2015,johnson2015,cao2016,banerjee2016}, which validates our method. More importantly, two pure meta-stable magnetic octupolar states $O^{36}_{21}$ (with FM and AFM configurations, respectively) without any magnetic dipoles are predicted. These two octupolar states can be stabilized by doping I elements in $\alpha$-\ce{RuCl3} or synthesizing $\alpha$-\ce{RuI3} directly, where the AFM octupolar state becomes the ground state. We propose that an orthogonal magnetization $\propto H^2\cos^2\theta$ can be detected as the fingerprint of the meta-stable FM octupolar state~\cite{okazaki2011rotational,tonegawa2012cyclotron,tian2017}.

\section{Model and Method}
In many $4d$ and $5d$ transition metal materials, the strong cubic crystal field  splits the five-fold $d$ orbitals into two-fold $e_g$ and three-fold $t_{2g}$ orbitals with electrons occupying only the low-energy $t_{2g}$ subspace, so a $t_{2g}$ model is sufficient for such systems~\cite{kim2008,kim2009,jackeli2009,pesin2010,witczak2014,rau2016}. For a $t_{2g}$ system, the local on-site Coulomb interaction can be well described by a multi-orbital Kanamori Hamiltonian~\cite{georges2013}, $H_{\text{int}}$. Under HFMF approximation, we can further express $H_{\text{int}}$ in terms of all the multipolar OPs in the $t_{2g}$ subspace as following (See Appendix~\ref{appa} for the details of derivations),

\begin{figure}[htbp]
    \centering
    \includegraphics[width=0.5\textwidth,trim=0 0 0 0,clip]{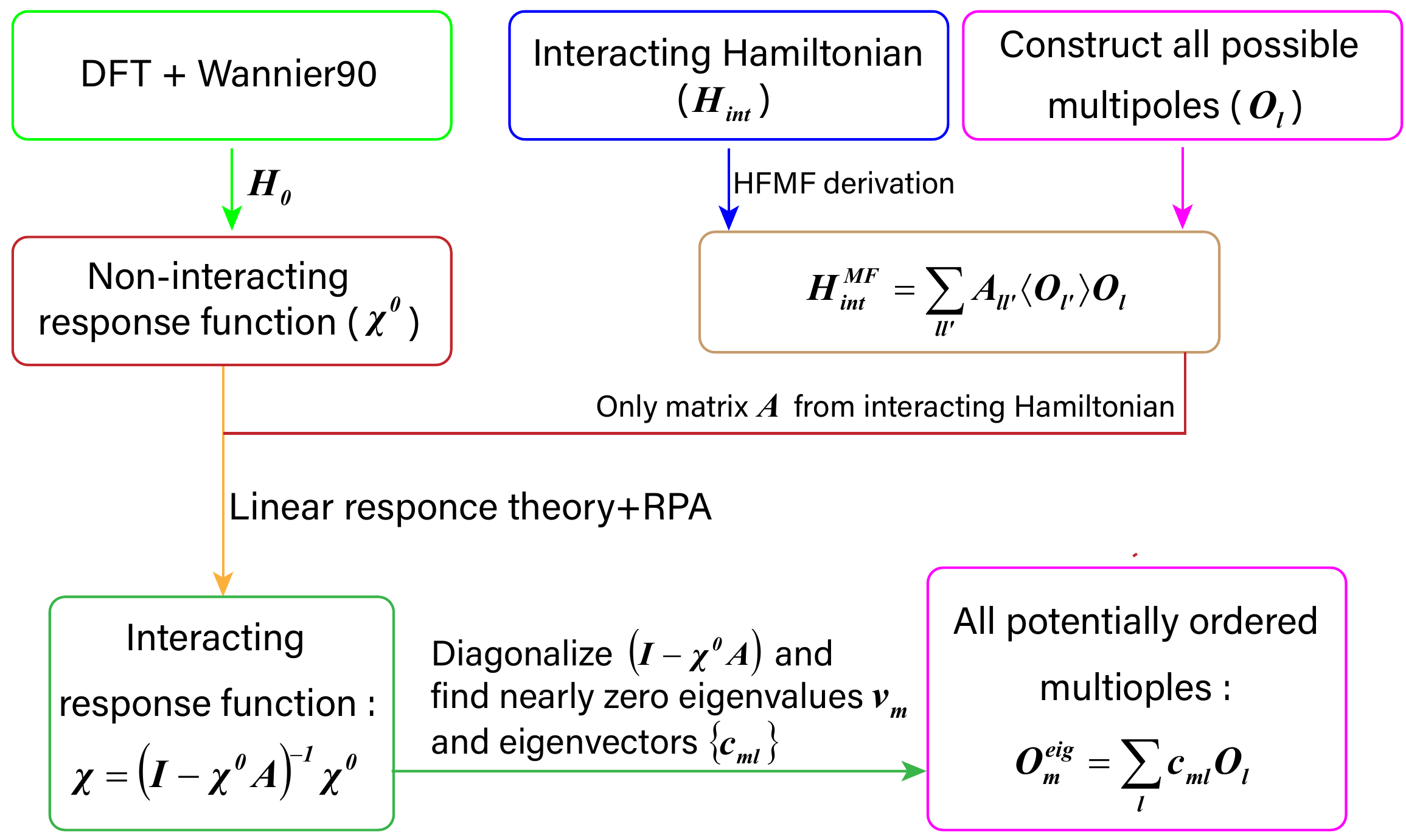}
    \caption{Flow diagram of predicting multipolar OPs based on DFT calculations and linear response theory with RPA.
    }
   \label{fig1}
    \end{figure}

\onecolumngrid
\begin{equation}
\label{mutipole_exp}
\begin{split}
H_{\text{int}}^{\text{MF}}=&(5U-10J)\langle{O}^{01}_{00}\rangle{O}^{01}_{00}
+(3J-U)\sum_{M=1}^{3}\langle {O}^{1M}_{10}\rangle{O}^{1M}_{10}
+(5J-U)\sum_{M=1}^{5}\langle {O}^{2M}_{20}\rangle{O}^{2M}_{20}
-(2J+U)\sum_{M=1}^{3}\langle {O}^{1M}_{01}\rangle{O}^{1M}_{01}\\
&+(3J-U)\left\{\langle{O}^{01}_{11}\rangle{O}^{01}_{11}
+\sum_{M=1}^{3}\langle {O}^{1M}_{11}\rangle{O}^{1M}_{11}
+\sum_{M=1}^{5}\langle {O}^{2M}_{11}\rangle{O}^{2M}_{11}\right\}\\
&+(J-U)\left\{\sum_{M=1}^{3}\langle{O}^{1M}_{21}\rangle{O}^{1M}_{21}+
\sum_{M=1}^{5}\langle{O}^{2M}_{21}\rangle{O}^{2M}_{21}+
\sum_{M=1}^{7}\langle{O}^{3M}_{21}\rangle{O}^{3M}_{21}\right\},
\end{split}
\end{equation}
\twocolumngrid
where, $U$ and $J$ are the Coulomb interaction and Hund’s coupling, respectively. ${O}^{KM}_{K_o K_s}$ are the 36 spin-orbital entangled multipoles with overall rank $K$ ($K=0\sim 3$ and $M=1,2,\cdots,2K+1$), which are composed of orbital and spin multipoles with rank $K_o$ and $K_s$~\cite{yilin2017}. ${O}^{01}_{00}$ and ${O}^{01}_{11}$ describe the charge and isotropic SOC terms, respectively. The electric quadrupoles ${O}^{2M}_{20}$ describe possible crystal field splitting in $t_{2g}$ subspace and ${O}^{2M}_{11}$ describe the corresponding anisotropic SOC effects. ${O}^{1M}_{10}$ and ${O}^{1M}_{01}$ are the conventional orbital and spin magnetic dipoles, and ${O}^{3M}_{21}$ describe magnetic octupoles. All the potentially ordered multipoles are contained in Eq.~\eqref{mutipole_exp}, from which we can intuitively capture their explicit physical implications.

Based on Eq.~\eqref{mutipole_exp} and the DFT constructed non-interacting Hamiltonian $H_0$, we use the LRT under RPA to determine which multipoles may actually occur. The flow diagram of our method is shown as Fig.~\ref{fig1}, in which self-consistent HFMF calculations are not needed. The basic formula of LRT can be written as
\begin{equation}
\label{mutipole_responce1}
\delta\langle{O}_{l}(\bm{q},\omega)\rangle=\sum_{l^{\prime}}\chi_{ll^{\prime}}(\bm{q},\omega)F^{ext}_{l^{\prime}}(\bm{q},\omega),
\end{equation}
where, $F^{ext}_{l^{\prime}}(\bm{q},\omega)$ is an external field coupled to a multipole $O_{l^{\prime}}$, and
$\chi_{ll^{\prime}}\propto \langle{[O_l,O_{l^{\prime}}]}\rangle_{\bm{q},\omega}$
is the interacting response function between multipoles $O_l$ and $O_{l^{\prime}}$. Under RPA, the local interactions in Eq.~\eqref{mutipole_exp} enter into $\chi$ only via a coefficient matrix $A$ composed of $U$ and $J$ (See the derivations of $A$ in Appendix~\ref{appa} and RPA in Appendix~\ref{appb}),
\begin{equation}
\label{mutipole_responce3}
\begin{split}
\chi=(I-{\chi}^{0} A)^{-1}{\chi}^{0},
\end{split}
\end{equation}
where, ${\chi}^{0}$ is the non-interacting response function obtained from the non-interacting Hamiltonian $H_0$, whose matrix element $\chi^{0}_{ll^{\prime}}$ is given by (See the derivations of $\chi^{0}_{ll^{\prime}}$ in Appendix~\ref{appc})
\begin{eqnarray}
\label{mutipole_responce2}
\chi^{0}_{ll^{\prime}}(\bm{q},\omega)&=&\frac{1}{\hbar N}\sum_{\alpha\beta\gamma\delta}O_l^{i_l \alpha\beta}
O_{l^{\prime}}^{i_{l^{\prime}} \delta\gamma}\Xi^{i_l i_{l^{\prime}}}_{\alpha\beta,\delta\gamma}(\bm{q},\omega),\\
\Xi^{i_l i_{l^{\prime}}}_{\alpha\beta,\delta\gamma}(\bm{q},\omega)&=&\sum_{\bm{k}jj^{\prime}} B^j_{i_{l^{\prime}}\gamma}(\bm{k}) B^{j*}_{i_{l}\alpha}(\bm{k})B^{j^{\prime}}_{i_{l}\beta}(\bm{k}+\bm{q}) B^{j^{\prime *}}_{i_{l^{\prime}}\delta}(\bm{k}+\bm{q}) \nonumber\\
&\times&\frac{f(\varepsilon_{j\bm{k}})-f(\varepsilon_{j^{\prime}\bm{k}+\bm{q}})}{\omega-(\varepsilon_{j^{\prime}\bm{k}+\bm{q}}-\varepsilon_{j\bm{k}})/\hbar+i0^{\dagger}}
\end{eqnarray}
where, $l$ labels the multipole $O_l$, $\alpha$, $\beta$, $\gamma$, $\delta$ label the spin-orbital basis, $j$ labels the Bloch band, $i_l$ labels the sub-lattice where $O_l$ resides, $B^j_{i_{l}\alpha}(\bm{k})$ is the $(i_{l},\alpha)$ component of the non-interacting wave-function of the $j$-th eigenstate at momentum $\bm{k}$ with eigenvalue $\varepsilon_{j\bm{k}}$, $f(\varepsilon_{j\bm{k}})$ is the Fermi distribution function. Since all the interacting effects only enter into  $A$, the interacting wave-functions are not required anymore and those time-consuming self-consistent HFMF calculations are avoided in our scheme, which leads to a very fast single-shot calculation.

\begin{figure}[t]
\centering
\includegraphics[width=0.5\textwidth,trim=0 0 0 0,clip]{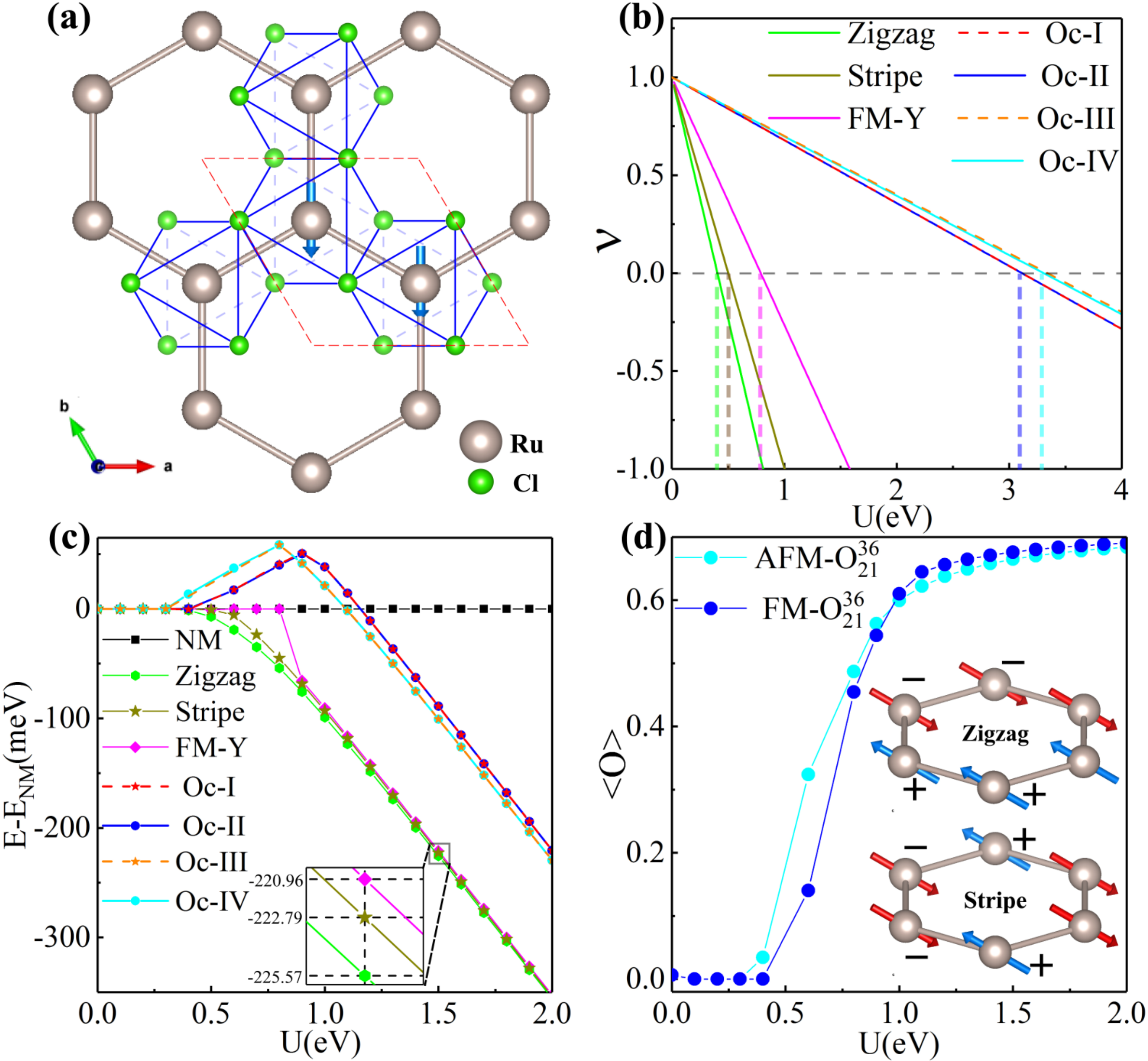}
\caption{(Color online). (a) Crystal structure of monolayer $\alpha$-\ce{RuCl3}. (b) The eigenvalues $\nu_{m}$ of $(I-{\chi}^{0} A)$ that approach to zero as a function of $U$. (c) Total energy of magnetic states relative to the NM state as a function of $U$. The inset shows that Zigzag magnetic order has the lowest energy. (d) The size of $O^{36}_{21}$ octupole moment. The insets show the Zigzag and Stripe antiferromagnetic orders respectively. $J=0$ eV and $\lambda=96$ meV are used in (b)-(d).
}
\label{fig2}
\end{figure}

Here, the divergence of $\chi$ is ambiguous in the original multipole representation of $\{O_l\}$ defined in Eq.~\eqref{mutipole_exp}, since the matrix $(I-\chi^0A)$ in the denominator of Eq.~\eqref{mutipole_responce3} is not diagonal. We can transform $\{O_l\}$ to a new (dubbed as eigen-order) representation  ${O}^{\textrm{eig}}_{m}=\sum_{l}c_{ml}{O}_{l}$ by diagonalizing $(I-\chi^0A)$, where $c_{ml}$ is the $l$-th component of the $m$-th eigenvector of $(I-{\chi}^{0}A)$. This indicates that the actually ordered parameter is usually a symmetry allowed combination of $\{O_{l}\}$.
Under RPA, the response matrix ${\chi}^{\prime}$ and ${{\chi}^{\prime}}^{0}$ in the new eigen-order representation satisfy the same relations as Eq.~\eqref{mutipole_responce3} and can be rewritten as ${\chi}^{\prime}_{mm^{\prime}}=\nu_{m}^{-1}{{\chi}^{\prime}}^{0}_{mm^{\prime}}$, where $\nu_{m}$ is the $m$-th eigenvalue of $(I-{\chi}^{0}A)$. Therefore, one can find spontaneous symmetry breaking and the corresponding OPs ${O}_{l}$ by checking the eigenvalues of $\nu_{m}$ that approach to zero and analyzing their eigenvectors $\{c_{m}\}$.

\begin{table}[t]
\newcommand{\tabincell}[2]{\begin{tabular}{@{}#1@{}}#2\end{tabular}}
\centering
\caption{The predicted ordered states, ${O}^{\textrm{eig}}_{m}=\sum_{l}c_{ml}{O}_{l}$, when $\nu_{m}\rightarrow0$ in monolayer $\alpha$-\ce{RuCl3}. The values in columns are the corresponding weights $c_{ml}$ of multipoles $O_l=O^{KM}_{K_o K_s}$. $+/-$ denote the moment directions at different Ru sites, as shown in Fig.~\ref{fig2} (d). Only the local OPs on the first Ru site are listed.}
\setlength{\tabcolsep}{0.15mm}{
\begin{tabular}{c|c|c|c|c|c|c|c}
\hline\hline
\diagbox{$O_l$}{$c_{ml}$}{${O}^{\text{eig}}_{m}$} & \tabincell{c}{Zigzag\\${+}{+}{-}{-}$} & \tabincell{c}{Stripe\\${+}{-}{-}{+}$} & \tabincell{c}{FM-Y\\${+}{+}$} & \tabincell{c}{Oc-I\\${+}{+}$} & \tabincell{c}{Oc-II\\${+}{+}$} & \tabincell{c}{Oc-III\\${+}{-}$} & \tabincell{c}{Oc-IV\\${+}{-}$}\\
\hline
$O^{11}_{10}$($l_x$) & -0.011 & 0.202 & 0.116 & 0 & 0 & 0 & 0\\
\hline
$O^{12}_{10}$($l_y$) & 0 & 0 & 0.065 & 0 & 0 & 0 & 0\\
\hline
$O^{13}_{10}$($l_z$) & -0.165 & -0.229 & 0 & 0.155 & 0 & -0.147 & 0\\
\hline
$O^{11}_{01}$($s_x$) & -0.058 & 0.121 & 0.594 & 0 & 0 & 0 & 0\\
\hline
$O^{12}_{01}$($s_y$) & 0 & 0 & 0.331 & 0 & 0 & 0 & 0\\
\hline
$O^{13}_{01}$($s_z$) & -0.052 & -0.140 & 0 & -0.043 & 0 & 0.257 & 0\\
\hline
$O^{33}_{21}$ & 0 & 0 & 0 & 0.652 & 0 & 0.626 & 0\\
\hline
$O^{36}_{21}$ & 0 & 0 & 0 & 0 & 0.707 & 0 & 0.691\\
\hline\hline
\end{tabular}}
\label{tab1}
\end{table}

\section{Application in monolayer $\alpha$-$\mathrm{RuCl_3}$}
Now we apply our method to search possible multipolar OPs in the monolayer $\alpha$-\ce{RuCl3} which crystallizes into a nearly ideal honeycomb lattice~\cite{plumb2014,sears2015,cao2016,banerjee2016,sandilands2016} with space group P-31m (No. 162), as shown in Fig.~\ref{fig2} (a). Similar to $\textrm{Ir}^\textrm{4+}$ in iridates, $\textrm{Ru}^\textrm{3+}$ with $d^5$ configuration will lead to half-filling of $j_\textrm{eff}=\frac{1}{2}$ states when SOC is considered~\cite{kim2008,kim2009,plumb2014,sandilands2016}. All the above features make \ce{RuCl3} a famous candidate of Kitaev spin liquid. Previous works on Kitaev physics in this material consider only the $j_\textrm{eff}=\frac{1}{2}$ states in the low-energy model~\cite{jackeli2009,chaloupka2010,banerjee2016,kim2016,leahy2017,baek2017,lampen2018,kasahara2018,hentrich2018}. However, comparing to its $5d$ counterparts such as \ce{Na2IrO3} and \ce{Li2IrO3}, relatively smaller SOC strength $\lambda$ in $4d$ $\textrm{Ru}^\textrm{3+}$ cannot effectively isolate the $j_\textrm{eff}=\frac{1}{2}$ from the $j_\textrm{eff}=\frac{3}{2}$ states~\cite{yilin2017} to induce a reasonable $j_\textrm{eff}=\frac{1}{2}$ single-orbital model ($\lambda=96$ meV for $\textrm{Ru}^\textrm{3+}$ and about $400$ meV for $\textrm{Ir}^\textrm{4+}$). Thus the multi-orbital degrees of freedom that are essential for multipolar OPs still play important roles here.

We first construct the non-interacting $t_{2g}$ tight-binding (TB) Hamiltonian $H_0$, based on the non-SOC DFT calculations by the Vienna \textit{ab initio} simulation package (VASP)~\cite{kresse1996} combined with the maximally localized Wannier functions method~\cite{arash2008,jan2010wien2wannier,ikeda2010phase,marzari2012}. The crystal symmetry of $H_0$ is restored using the code developed by Yue~\cite{symm}, whose band structures match well with the DFT bands (See Fig.~\ref{figs1} in Appendix~\ref{appd}). An atomic SOC term of $\lambda\bm{l}\cdot\bm{s}$ with $\lambda=96$ meV from the optical spectroscopy experiment~\cite{sandilands2016} is added to $H_0$. The non-interacting response matrix $\chi_0$ is calculated according to Eq.~\eqref{mutipole_responce2} using the eigen-energy and wave-functions of $H_0+\lambda\bm{l}\cdot\bm{s}$. We then diagonalize $(I-{\chi}^{0}A)$ to obtain its eigenvalues $\nu_m$ and the corresponding eigenvectors $\{c_{m}\}$. In Fig.~\ref{fig2} (b), we plot $\nu_{m}$ that approach to zero as a function of $U$. The first one approaching to zero is the green curve at $U=0.4$ eV, which corresponds to the most likely occurred order in $\alpha$-\ce{RuCl3}. By analyzing $c_{ml}$ as shown in the 2nd column of Table~\ref{tab1}, we find that they are magnetic dipoles, ${O}^{11}_{10}$ (${O}^{11}_{01}$) and ${O}^{13}_{10}$ (${O}^{13}_{01}$) with $\bm{q}=[0,0.5]$, corresponding to the Zigzag configuration [see the insert of Fig.~\ref{fig2} (d)]. This is consistent with the neutron scattering experiment~\cite{cao2016} and thus validates our method. The second (dark yellow) and third (pink) divergent terms correspond to the Stripe ($\bm{q}=[0,0.5]$) and FM-Y [Fig.~\ref{fig2} (a)] magnetic orders, respectively, which are also widely studied for $\alpha$-\ce{RuCl3}~\cite{kim2015,hou2017}.

Besides these extensively studied magnetic dipolar states, we find that monolayer $\alpha$-\ce{RuCl3} may also enter four new magnetic octupolar states (Oc-I$\sim$IV). As shown in Table \ref{tab1}, Oc-I (Oc-III) state is dominated by the FM (AFM) arranged magnetic octupole $O^{33}_{21}$ accompanying with minor magnetic dipole components, while Oc-II (Oc-IV) state has a pure magnetic octupole moment $O^{36}_{21}$ with FM (AFM) arrangement. We notice that $O^{33}_{21}$ is the counterpart of $O^{36}_{21}$ by an operation of $x\leftrightarrow -y$, according to the original definitions~\cite{yilin2017}. Their difference is that $O^{36}_{21}$ respects all the point group symmetries (including $\textrm{C}_\textrm{3z}$ and $\textrm{C}_\textrm{2y}$) of the non-interacting Hamiltonian $H_0$ of monolayer $\alpha$-\ce{RuCl3}, while $O^{33}_{21}$ respects  $\textrm{C}_\textrm{3z}$ but not $\textrm{C}_\textrm{2y}$ symmetry. Therefore, $O^{33}_{21}$ can coexist with the magnetic dipoles that align along the $z$-direction (more analyses are given in Appendix~\ref{appf}).

We also perform self-consistent unrestricted HFMF calculations to check the above predictions from our new method. To drive the system to a desired ordered state, we symmetrize the mean-field Hamiltonian according to its magnetic group at each step of iterations. The total energy of the converged ordered states relative to the non-magnetic (NM) state as functions of $U$ are plotted in Fig.~\ref{fig2} (c), which confirms our prediction that the Zigzag antiferromagnetic ordered state is the ground state when $U>0.4$ eV, whose energy is about $2\sim4$ meV lower than the other two magnetic dipolar states. We notice that the Oc-I$\sim$IV states can also be stabilized by $U$ around $0.5\sim1.1$ eV ($J=0$ eV) although their energy are higher than NM state. When $U$ exceeds $1.15$ eV, their energy become lower than the NM state. The energy difference between Oc-I (Oc-III) and Oc-II (Oc-IV) is very tiny at $J=0$ eV (Oc-II is about $10^{-4}$ meV lower than Oc-I). When $J\ge0.22$ eV, the Oc-I (Oc-III) state can not be stabilized anymore (Appendix~\ref{appf}). Therefore, we only study Oc-II (Oc-IV) state, i.e. FM-$O^{36}_{21}$ (AFM-$O^{36}_{21}$) state, hereafter.

The calculated size of the $O^{36}_{21}$ octupole moments in FM- and AFM-$O^{36}_{21}$ states with respect to $U$ are plotted in Fig.~\ref{fig2} (d). It shows that this octupole moment appears around $U=0.5$ eV, then increases monotonously as increasing $U$, and finally saturates once entering into the meta-stable state. These features show a typical first-order phase transition~\cite{khomskii2010} from NM to the $O^{36}_{21}$ state. All the self-consistent HFMF calculations are consistent with our predictions, which validates our new method.

\begin{figure}[t]
\centering
\includegraphics[width=0.5\textwidth,trim=0 0 0 0,clip]{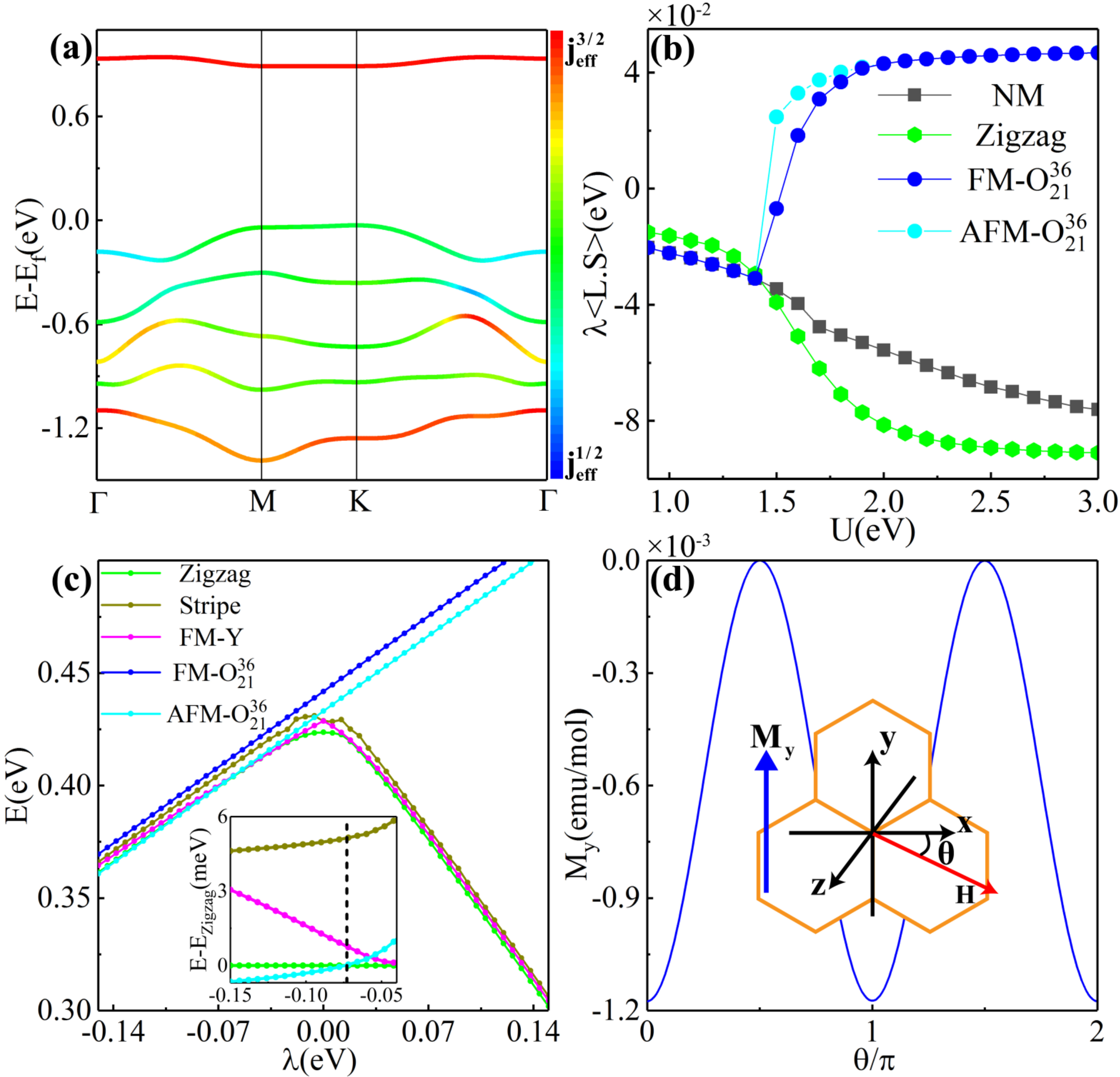}
\caption{(Color online). (a) Band structures of the AFM-$O^{36}_{21}$ state at the experimental $U=2.4$ eV and $\lambda=96$ meV. The color-bar shows orbital projection of $j_{\text{eff}}=1/2,3/2$. (b) Effective SOC $\lambda\langle\bm{l}\cdot\bm{s}\rangle$ of NM, Zigzag, FM- and AFM-$O^{36}_{21}$ states as a function of $U$ at $\lambda=96$ meV.
(c) Energy of magnetic states of monolayer $\alpha$-\ce{RuCl3} as a function of $\lambda$ at $U=2.4$ eV.
(d) Orthogonal magnetic moment $M_y$ produced by $O^{36}_{21}$ octupole calculated at $U=2.4$ eV and $\lambda=96$ meV under a rotating magnetic field of 5 T applied in the $xz$ plane. $J=0.4$ eV is used in (a)-(d).
}
\label{fig3}
\end{figure}

We now study the electronic structures of the $O^{36}_{21}$ states. Here, in Fig.~\ref{fig3} (a) we plot the band structures of AFM-$O^{36}_{21}$ state since it is more favorable in energy [about 12 meV lower than FM-$O^{36}_{21}$ state, see Fig.~\ref{fig2} (c)]. The color-bar in Fig.~\ref{fig3} (a) shows orbital projection of $j_\textrm{eff}=\frac{1}{2}, \frac{3}{2}$. Different from the typical band structures of a $t_{2g}$ system with SOC, the unoccupied bands in AFM-$O^{36}_{21}$ state are mainly $j_\textrm{eff}=\frac{3}{2}$ type rather than $j_\textrm{eff}=\frac{1}{2}$ type, which is caused by an interaction-induced positive SOC $\lambda\langle\bm{l}\cdot\bm{s}\rangle$ [see cyan curve in Fig.~\ref{fig3} (b)]. On the contrary, the Zigzag and NM states have negative $\lambda\langle\bm{l}\cdot\bm{s}\rangle$.

In the following, we would like to discuss how to stabilize such $O^{36}_{21}$ states in materials. (1) Two rotation symmetries, which can forbid the presence of dipolar OPs, are the necessary condition to protect such pure octupole. (2) Our results in Fig.~\ref{fig3} (b) obviously demonstrate that negative SOC ($-{\lambda}$) could reduce the energy of $O^{36}_{21}$ states by $\lambda \langle\bm{l}\cdot \bm{s}\rangle$.
In Fig.~\ref{fig3} (c), we plot the energy versus $\lambda$ for different magnetic states in monolayer $\alpha$-\ce{RuCl3} at experimental $U$ and $J$, which indicates that negative $\lambda$ indeed makes the energy of $O^{36}_{21}$ states much closer to dipolar states. More interestingly, when $\lambda<-73$ meV, the AFM-$O^{36}_{21}$ state becomes the ground state, whose energy is $\sim1$ meV lower than Zigzag dipolar ordered state. In real materials, this can be achieved by mixing $t_{2g}$ orbitals with more $p$ orbitals, since $-\lambda \bm{l}_{t_{2g}} \cdot \bm{s} = \lambda (-\bm{l}_{t_{2g}})\cdot \bm{s}=\lambda \bm{l}_{p}\cdot \bm{s}$~\cite{sheng2014,nie2017}. This can be realized by doping I elements in $\alpha$-\ce{RuCl3} or synthesizing $\alpha$-\ce{RuI3} directly. As shown in Fig.~\ref{figs4}, the orbital projection of the hypothetical monolayer $\alpha$-\ce{RuI3} with an optimized structure exhibits an enhancement of the $j_{\text{eff}}=3/2$ character around the Fermi level compared to $\alpha$-\ce{RuCl3}, which implies that a negative $\lambda\approx$ -100 meV is realized. We then calculate the magnetic states of $\alpha$-\ce{RuI3} using the same method, as shown in Fig.~\ref{figs5}, where the AFM-$O^{36}_{21}$ state becomes the ground state with its energy about 3 meV lower than the Zigzag dipolar state.

Finally, we would like to discuss how to detect the octupolar states $O^{36}_{21}$. Under an external magnetic field $H$, the free energy contributed by $O^{36}_{21}$ is proportional to $3H^2_x H_y-H^3_y$ (Appendix~\ref{appf}). Its $H_y$ derivative gives rise to a magnetic moment in the $y$ direction as the form of $M_y\propto H^2_x-H^2_y$. Thus, an orthogonal magnetization oscillation of $M_y\propto H^2\cos^2\theta$ would be expected if a rotating magnetic field $H$ is applied in the $xz$ plane with $\theta$ respect to $x$-axis.
Fig.~\ref{fig3} (d) shows the calculated $M_y$ of the FM-$O^{36}_{21}$ state in monolayer $\alpha$-\ce{RuCl3} as a function of $\theta$ under a magnetic field of 5 T. The induced $M_y$ is about ~$10^{-3}$ emu/mol, which is completely contributed by the octupole $O^{36}_{21}$ since no dipole exists, in contrast to the case in \ce{Eu2Ir2O7}~\cite{tian2017,yilin2017}. This can be taken as a fingerprint for experimental detection of $O^{36}_{21}$ state. However, no orthogonal magnetization could be detected in AFM-$O^{36}_{21}$ state since the induced $M_y$ on two $\textrm{Ru}^\textrm{3+}$ cancel out. How to detect the AFM-$O^{36}_{21}$ state is an open question and requires further study.

\section{Conclusion}
In summary, we have presented an efficient method to predict meta-stable/ground multipolar states in real materials, in which both electronic correlation and SOC play important roles. We apply this method to study $\alpha$-\ce{RuX3} (X=Cl,I). It has not only correctly reproduced the magnetic ground state observed in experiments, but also successfully predicted two meta-stable magnetic octupolar states in $\alpha$-\ce{RuCl3}, which are confirmed by further self-consistent unrestricted HFMF calculations.
We show that these meta-stable magnetic octupolar states can be stabilized and the AFM-$O^{36}_{21}$ even becomes the ground state in $\alpha$-\ce{RuI3} via mixing $t_{2g}$ orbitals with more $p$ components.
We also predict that an orthogonal magnetization $M_y$ can arise from the FM-$O^{36}_{21}$ state, which is the fingerprint and can be easily detected by magnetic torque experiment~\cite{okazaki2011rotational,tonegawa2012cyclotron,tian2017}. Our scheme serves as a guidance for efficient prediction and realization of meta-stable/ground multipolar states in $d$-orbital systems.

\section*{Acknowledgments}
The authors thank Xi Dai for helpful discussions. The authors acknowledge
the support by the National Key Research and Development
Program of China (2018YFA0307000,2017YFA0403501), and the National Natural Science Foundation of China (11874022,11874160). Yi-Lin Wang is supported by USTC Research Funds of the Double First-Class Initiative (No.~YD2340002005). Jin-Yu Zou is supported by the China Postdoctoral Science Foundation (2019M662580).

\appendix
\onecolumngrid

\renewcommand\thefigure{\Alph{section}\arabic{figure}}
\renewcommand\theequation{\Alph{section}\arabic{equation}}

\section{MULTIPOLE EXPANSION OF LOCAL COULOMB INTERACTION HAMILTONIAN}\label{appa}
\setcounter{figure}{0}
\setcounter{equation}{0}

The local Coulomb interaction of $t_{2g}$ orbitals is well-described by a multi-orbital Kanamori Hamiltonian~\cite{georges2013} and it reads,
\begin{equation}
\begin{aligned}
H_{\text{int}}=& U \sum_{m} \hat{n}_{m \uparrow} \hat{n}_{m \downarrow}+U^{\prime} \sum_{m \neq m^{\prime}} \hat{n}_{m \uparrow} \hat{n}_{m^{\prime} \downarrow}+\left(U^{\prime}-J\right) \sum_{m<m^{\prime}, s} \hat{n}_{m s} \hat{n}_{m^{\prime} s}\\
&-J \sum_{m \neq m^{\prime}} d_{m \uparrow}^{\dagger} d_{m \downarrow} d_{m^{\prime} \downarrow}^{\dagger} d_{m^{\prime} \uparrow}+J \sum_{m \neq m^{\prime}} d_{m \uparrow}^{\dagger} d_{m \downarrow}^{\dagger} d_{m^{\prime} \downarrow} d_{m^{\prime} \uparrow},
\end{aligned}
\end{equation}
where, $\hat{n}_{m s}=d_{m s}^{\dagger}d_{m s}$ is the electron number operator with orbital $m$ and spin $s$ ($\uparrow$ or $\downarrow$). For convenience, we use $\alpha$, $\beta$, $\gamma$, $\delta$ to label the spin-orbital coupled basis of $\{m s$\}. In general, $H_{\text{int}}$ can be represented as
\begin{equation}
H_{\text{int}}=\sum_{\alpha<\beta, \gamma<\delta} U_{\alpha \beta, \gamma \delta} d_{\alpha}^{\dagger} d_{\beta}^{\dagger} d_{\delta} d_{\gamma}.
\end{equation}
Under the unrestricted Hartree-Fock mean-field (HFMF) approximation, it can be written in a single particle form,
\begin{equation}\label{HpartMF}
\begin{aligned}
H^{\mathrm{MF}}_{\mathrm{int}}=&\sum_{\alpha<\beta, \gamma<\delta} U_{\alpha \beta, \gamma \delta}\left(n_{\alpha \gamma} d_{\beta}^{\dagger} d_{\delta}+n_{\beta \delta} d_{\alpha}^{\dagger} d_{\gamma}-n_{\alpha \delta} d_{\beta}^{\dagger} d_{\gamma}-n_{\beta \gamma} d_{\alpha}^{\dagger} d_{\delta}-n_{\alpha \gamma} n_{\beta \delta}+n_{\alpha \delta} n_{\beta \gamma}\right) \\
=&\sum_{\beta \delta} V_{\beta \delta} d_{\beta}^{\dagger} d_{\delta}+\sum_{\alpha<\beta, \gamma<\delta}\left(n_{\alpha \delta} n_{\beta \gamma}-n_{\alpha \gamma} n_{\beta \delta}\right),
\end{aligned}
\end{equation}
where, $n_{\alpha \gamma}$ is the local density matrix $\left\langle d_{\alpha}^{\dagger} d_{\gamma}\right\rangle$, the last term contributes an energy constant, and $V_{\beta \delta}$ reads
\begin{equation}\label{HpartMF_element}
\begin{aligned}
V_{\beta \delta}=\left(\sum_{\alpha<\beta, \gamma<\delta} U_{\alpha \beta, \gamma \delta}
+\sum_{\alpha>\beta, \gamma>\delta} U_{\beta \alpha, \delta \gamma}
-\sum_{\alpha<\beta, \gamma>\delta} U_{\alpha \beta, \delta \gamma}
-\sum_{\alpha>\beta, \gamma<\delta} U_{\beta \alpha, \gamma \delta}\right) n_{\alpha \gamma}
=\sum_{\alpha \gamma} A_{\alpha \gamma}^{\beta \delta} n_{\alpha \gamma}.
\end{aligned}
\end{equation}
Now we expand Eq.~\eqref{HpartMF} in terms of the multipoles $O_{l}$, namely
\begin{equation}\label{HpartMF2}
H^{\mathrm{MF}}_{\mathrm{int}}=\sum_{l} V_{l} O_{l}, \quad
O_{l}=\sum_{\beta \delta} O_{l}^{\beta\delta} d_{\beta}^{\dagger} d_{\delta},
\end{equation}
where, $O_{l}^{\beta\delta}$ is the matrix element of $O_{l}$. According to Ref.~\cite{yilin2017}, $O_{l}$ are orthogonal and complete,
\begin{equation}\label{trace}
\sum_{\beta\delta}O_{l}^{\beta\delta} [O_{l^{\prime}}^{\dagger}]^{\delta\beta}=\delta_{l l^{\prime}},\quad
\sum_{l}O_{l}^{\beta\delta} [O_{l}^{\dagger}]^{\alpha\gamma}=\delta_{\beta\alpha}\delta_{\delta\gamma}.
\end{equation}
Then we can inversely express the second equation of Eq.~\eqref{HpartMF2} as
\begin{equation}\label{ldm}
d_{\alpha}^{\dagger} d_{\gamma}=\sum_{l} O_{l} [O_{l}^{\gamma\alpha}]^{*}, \quad
n_{\alpha \gamma}=\sum_{l} \left\langle O_{l} \right\rangle [O_{l}^{\gamma\alpha}]^{*}.
\end{equation}
Combining Eq.~\eqref{HpartMF}, Eq.~\eqref{HpartMF_element}, Eq.~\eqref{HpartMF2}, Eq.~\eqref{ldm} and using Eq.~\eqref{trace} again, we obtain
\begin{equation}\label{mtxA}
V_{l}=\sum_{l^{\prime}} A_{l l^{\prime}}\left\langle O_{l^{\prime}}\right\rangle, \quad
A_{l l^{\prime}}=\sum_{\alpha \gamma ; \beta \delta} [O_{l}^{\beta \delta}]^{*} A_{\alpha \gamma}^{\beta \delta} [O_{l^{\prime}}^{\gamma\alpha}]^{*}.
\end{equation}
By substituting Eq.~\eqref{mtxA} into Eq.~\eqref{HpartMF2} we get Eq.~\eqref{mutipole_exp} in our main text, where $A$ is the coefficient matrix composed of $U$ and $J$, as shown in Eq.~\eqref{mutipole_exp} and Fig.~\ref{fig1}.

Note that the original form of mean-field Hamiltonian [Eq.~\eqref{HpartMF}] is expressed in terms of local density matrix $\left\langle d_{\alpha}^{\dagger} d_{\gamma}\right\rangle$, which is ambiguous to understand the effect of $U$, $J$ and $\lambda$ on multipoles and their interplays. The explicit expression in Eq.~\eqref{mutipole_exp} is more physical than Eq.~\eqref{HpartMF} since we can intuitively capture the physical implications. For example, the term $(3J-U)\left\langle O_{11}^{01}\right\rangle O_{11}^{01}=(J-U/3)\langle\bm{l} \cdot \bm{s}\rangle \bm{l}\cdot\bm{s}$ in Eq.~\eqref{mutipole_exp} can be regarded as an interaction-induced effective SOC. Therefore, the total SOC term of an interacting system can be written as ${\lambda}_{\mathrm{tot}}=\lambda+{\lambda}_{\mathrm{int}}$, ${\lambda}_{\mathrm{int}}=(J-U/3)\langle\bm{l} \cdot \bm{s}\rangle$. It can either enhance the SOC effect with ${\lambda}_{\mathrm{int}}>0$, or reduce the SOC effect with ${\lambda}_{\mathrm{int}}<0$ and even change the sign of ${\lambda}_{\mathrm{tot}}$.


\section{LINEAR RESPONSE THEORY UNDER RANDOM PHASE APPROXIMATION}\label{appb}
\setcounter{figure}{0}
\setcounter{equation}{0}

We assume an external field $F^{\mathrm{ext}}_{l}$ that couples to a multipolar operator $O_{l}$. The perturbation Hamiltonian is given by~\cite{jishi2013}
\begin{equation}\label{perturbation1}
\delta H^{\mathrm{ext}}=\sum_{l} F^{\mathrm{ext}}_{l} O_{l}.
\end{equation}
We first consider the non-interacting case. According to linear response theory (LRT), the variation of the ensemble average $\left\langle O_{l} \right\rangle$ induced by this perturbation is described by
\begin{equation}\label{hfrsp}
\delta\left\langle {O}_{l} \right\rangle^{\mathrm{ext}}=\sum_{l^{\prime}} \chi_{l l^{\prime}}^{0} F^{\mathrm{ext}}_{l^{\prime}},
\end{equation}
where, $\chi_{ll^{\prime}}^{0}\propto \langle{[O_l,O_{l^{\prime}}]}\rangle_0$ is the correlation function between multiples $O_{l}$ and $O_{l^{\prime}}$ with their commutator represented by the square bracket. The superscript ``0" denotes that the response function is calculated from the non-interacting Hamiltonian $H_0$.

Then we consider the case where the local Coulomb interacting effects of $H_\text{int}$ exist. Since its HFMF expression $H^{\mathrm{MF}}_{\mathrm{int}}$ is given by Eq.~\eqref{HpartMF2} and Eq.~\eqref{mtxA}, $\delta\left\langle {O}_{l} \right\rangle$ will induce an additional perturbation Hamiltonian, which reads,
\begin{equation}\label{perturbation2}
\delta H^{\mathrm{ind}}=\delta H^{\text{MF}}_{\text{int}}=\sum_{l l^{\prime}} A_{l l^{\prime}} \delta \left\langle O_{l^{\prime}}\right\rangle O_{l}.
\end{equation}
Combining Eq.~\eqref{perturbation1} and Eq.~\eqref{perturbation2}, the total field that couples $O_{l}$ can be expressed as
\begin{equation}
F_{l}=F^{\mathrm{ext}}_{l}+F^{\mathrm{ind}}_{l}, \quad
F^{\mathrm{ind}}=\sum_{l^{\prime}} A_{l l^{\prime}} \delta \left\langle O_{l^{\prime}}\right\rangle, \quad
\delta\left\langle {O}_{l^{\prime}} \right\rangle=\sum_{l^{\prime\prime}} \chi_{l^{\prime}l^{\prime\prime}} F^{\mathrm{ext}}_{l^{\prime\prime}}.
\end{equation}
Correspondingly, the total variation of $\left\langle O_{l} \right\rangle$ induced by this total field can be written as a matrix form,
\begin{equation}
\delta{\left\langle O \right\rangle}=\chi^{0} F=\chi^{0} F^{\mathrm{ext}}+\chi^{0} A \chi F^{\mathrm{ext}}=\chi F^{\mathrm{ext}},
\end{equation}
where in the first step, interacting system perturbed by external field $F^{\mathrm{ext}}$ is treated as equivalent to non-interacting system perturbed by total field $F$. This is called the random phase approximation (RPA). In the last step, the effect of $F^{\mathrm{ind}}$ induced by $H^{\text{MF}}_{\text{int}}$ has entered into the interacting response function $\chi$ only via a coefficient matrix $A$, which reads,
\begin{equation}\label{rpaform}
\chi=(I-\chi^{0} A)^{-1} \chi^{0}.
\end{equation}

Actually, $\chi$ is just the interacting response function got by Green's function method with RPA~\cite{jishi2013}.


\section{Derivations of non-interacting response function ${\chi}^{0}$}\label{appc}
\setcounter{figure}{0}
\setcounter{equation}{0}

We now give the derivations of non-interacting response function ${\chi}^{0}$. Its expression after the Fourier transformation with respect to space and time is given by~\cite{jishi2013},
\begin{equation}
\chi_{l l^{\prime}}^{0}\left(\bm{q}, \omega\right)=-\frac{1}{\hbar N}\int dt\left\{i\theta\left(t\right)\left\langle\left[{O}_{l}\left(\bm{q}, t\right), {O}_{l^{\prime}}\left(-\bm{q}, 0\right)\right]\right\rangle\right\} e^{i \omega t},
\end{equation}
where, ${O}_l(\bm{q}, t)=\sum_{\alpha \beta, \bm{k}} O^{i_l \alpha \beta}_l d_{i_l \alpha; \bm{k}}^{\dagger}(t) d_{i_l \beta; \bm{k}+\bm{q}}(t)$, $i_l$ labels the sub-lattice where $O_l$ resides. For convenience of derivations, we first calculate the imaginary-time ($\tau$) correlation function,
\begin{equation}\label{imagersp}
\begin{aligned}
X_{l l^{\prime}}^{0}\left(\bm{q},\tau\right)
=\frac{1}{\hbar N} \sum_{\alpha \beta \gamma \delta} O^{i_l \alpha \beta}_{l} O^{i_{l^{\prime}} \delta \gamma}_{l^{\prime}} \sum_{\bm{k}} g_{i_{l^{\prime}}i_l}^{0}\left(\gamma \alpha,\bm{k},-\tau\right) g_{i_l i_{l^{\prime}}}^{0}(\beta \delta,\bm{k}+\bm{q},\tau),
\end{aligned}
\end{equation}
where, $g^0$ is the non-interacting Green's function. In the eigenstate representation, it can be written as
\begin{equation}\label{green1}
g_{i_{l^{\prime}} i_l}^{0}(\gamma \alpha,\bm{k},-\tau)=\sum_{j} B_{i_{l^{\prime}} \gamma}^{j}(\bm{k}) B_{i_{l} \alpha}^{j*}(\bm{k}) g^{0}(j \bm{k},-\tau),
\end{equation}
where, $B^j_{i_{l}\alpha}(\bm{k})$ is the $(i_{l},\alpha)$ component of the non-interacting wave-function of the $j$-th eigenstate at momentum $\bm{k}$ with eigenvalue $\varepsilon_{j\bm{k}}$. The Fourier transformation of $X_{l l^{\prime}}^{0}\left(\bm{q},\tau\right)$ and $g^{0}(j \bm{k},\tau)$ with respect to $\tau$ are expressed as
\begin{equation}\label{green2}
X_{l l^{\prime}}^{0}\left(\bm{q},\tau\right)=\frac{1}{\beta\hbar}\sum_{n} X_{l l^{\prime}}^{0}\left(\bm{q},\omega_n\right) e^{-i \omega_n \tau}, \quad
g^{0}(j \bm{k},\tau)=\frac{1}{\beta\hbar}\sum_{n} g^{0}(j \bm{k},\omega_n) e^{-i \omega_n \tau},
\end{equation}
where $g^{0}(j \bm{k},\omega_n)$ is given by $(i\omega_n-\varepsilon_{j\bm{k}})^{-1}$. Combining Eq.~\eqref{imagersp}, Eq.~\eqref{green1}, Eq.~\eqref{green2} and using ${\chi}^{0}(\bm{q}, \omega)=X^{0}\left(\bm{q}, i \omega \rightarrow \omega+i 0^{\dagger}\right)$, we obtain
\begin{equation}
\begin{aligned}
\chi^{0}_{ll^{\prime}}(\bm{q},\omega)=&\frac{1}{\hbar N}\sum_{\alpha\beta\gamma\delta}O_l^{i_l \alpha\beta}
O_{l^{\prime}}^{i_{l^{\prime}} \delta\gamma}\Xi^{i_l i_{l^{\prime}}}_{\alpha\beta,\delta\gamma},\\
\Xi^{i_l i_{l^{\prime}}}_{\alpha\beta,\delta\gamma}=\sum_{\bm{k}jj^{\prime}} B^j_{i_{l^{\prime}}\gamma}(\bm{k}) B^{j*}_{i_{l}\alpha}(\bm{k})&B^{j^{\prime}}_{i_{l}\beta}(\bm{k}+\bm{q}) B^{j^{\prime *}}_{i_{l^{\prime}}\delta}(\bm{k}+\bm{q})
\times\frac{f(\varepsilon_{j\bm{k}})-f(\varepsilon_{j^{\prime}\bm{k}+\bm{q}})}{\omega-(\varepsilon_{j^{\prime}\bm{k}+\bm{q}}-\varepsilon_{j\bm{k}})/\hbar+i0^{\dagger}}.
\end{aligned}
\end{equation}

\section{Details of DFT calculations}\label{appd}
\setcounter{figure}{0}
\setcounter{equation}{0}

\begin{figure}[t]
\centering
\includegraphics[width=0.8\textwidth,trim=0 0 0 0,clip]{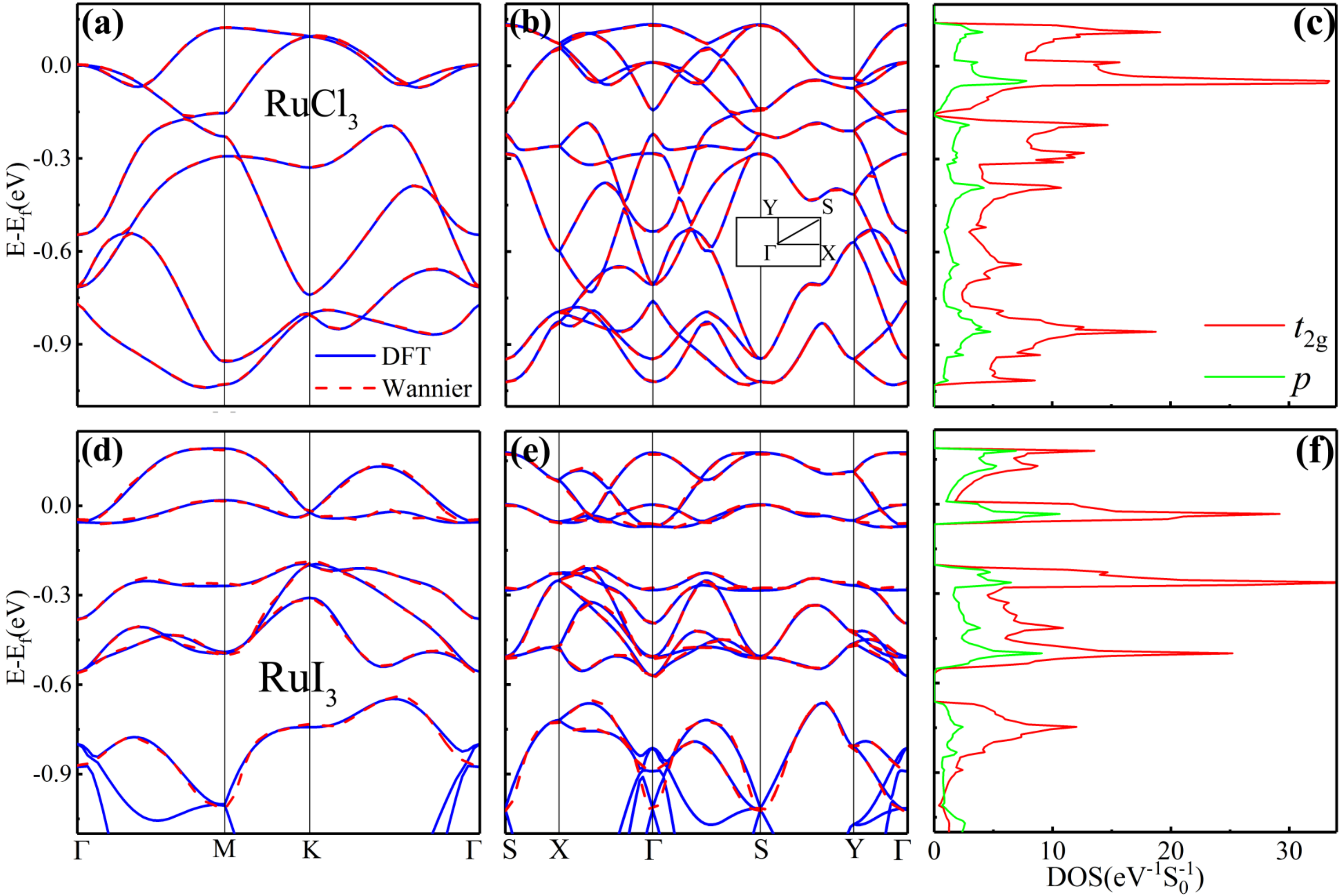}
\caption{(Color online). Non-interacting band structures and DFT calculated projected density of states without SOC included. Blue lines are the bands obtained by DFT calculations, red dashed lines are the bands obtained by Wannier functions method. (a) Bands of $\alpha$-\ce{RuCl3} with a primitive cell. (b) Bands of $\alpha$-\ce{RuCl3} with a supercell. (c) Density of states of $\alpha$-\ce{RuCl3}. (d) Bands of $\alpha$-\ce{RuI3} with a primitive cell. (e) Bands of $\alpha$-\ce{RuI3} with a supercell. (f) Density of states of $\alpha$-\ce{RuI3}.
}
\label{figs1}
\end{figure}

The DFT calculations are performed by the Vienna Ab-initio Simulation Package (VASP), where projector-augmented wave method and a plane wave basis set are used~\cite{kresse1996}. We select the Perdew-Burke-Ernzerhof (PBE) version of the generalized gradient approximation~\cite{PBE}. The plane wave cutoff energy is $600$ eV. We sample the Brillouin zone by $\Gamma$ centered scheme with a $9\times9\times3$ $k$-point mesh. The crystal structure of the monolayer $\alpha$-\ce{RuCl3} is displayed in Fig.~\ref{fig2} (a), where the lattice constant $a_0=5.8$ $\AA$ and the height of $\textrm{Cl}^\textrm{-}$ $h_\mathrm{Cl}=1.4$ $\AA$. The crystal structures of monolayer $\alpha$-\ce{RuI3} is obtained by an optimizing calculation (lattice constant $a_0=6.7$ $\AA$ and the height of $\textrm{I}^\textrm{-}$ $h_\mathrm{I}=1.5$ $\AA$), where the convergence criteria for force acting on each atom is set to $<10^{-2}$ eV/$\AA$. To avoid the interaction between the nearest layers, we set the inter-layer vacuum space to be $17$ $\AA$.

We first obtain the energy bands based on the non-SOC DFT calculations without $U$. The results are shown in Fig.~\ref{figs1} (a) and (d) for $\alpha$-\ce{RuCl3} and $\alpha$-\ce{RuI3}, respectively. Then we use maximally localized Wannier functions method~\cite{arash2008,jan2010wien2wannier,ikeda2010phase,marzari2012} to construct the non-interacting $t_{2g}$ tight-binding (TB) Hamiltonian $H_0$. We restore the symmetry of $H_0$ using the \href{https://github.com/yuechm/Wannier_Hamiltonian_symmetrization}{code developed by Yue} and then obtain the wannier bands by diagonalizing $H_0$, which can match well with the DFT bands, as shown in Fig.~\ref{figs1} (a) and (d). To perform the RPA calculations of Zigzag and Stripe orders, we transform the $H_0$ of the primitive cell into the one of a supercell, whose lattice vector is enlarged by two times in direction $\bm{b}$. The corresponding band structures are shown in Fig.~\ref{figs1} (b) and (e).

After constructing the non-interacting Hamiltonian $H_0$, we perform RPA and self-consistent calculations based on $H_0+H_{\text{int}}$. As mentioned in the main text, we only focus on the $t_{2g}$ model, rather than a $d-p$ model. Therefore, the double-counting term is just a constant for specific Coulomb interaction $U$ and can be absorbed into the chemical potential, as pointed out in Ref.~\cite{yilin2017}. When we compare the total energy between different magnetic orders at the same Coulomb interaction $U$, the double-counting term is a same constant due to the same local occupation number, \textit{e.g.}~that is 5 in $\alpha$-\ce{RuCl3}. Therefore, the double-counting will not cause any problem in our calculations.

\section{Details of RPA calculations}\label{appe}
\setcounter{figure}{0}
\setcounter{equation}{0}

In our RPA calculations of Fig.~\ref{fig2} (b), the FM-Y dipolar state and four octupolar states are identified based on $\bm{q}=[0,0]$ calculations, while the Zigzag and Stripe dipolar states correspond to the calculations with $\bm{q}=[0,0.5]$ of the primitive cell. Here, we would like to explain that our non-zero $\bm{q}$ calculations are realized by constructing the corresponding supercell based on $H_0$. For example, the Zigzag and Stripe ordered states in Fig.~\ref{fig2} (b) are calculated based on a $\bm{b}$-direction doubled supercell with $\bm{q}_\text{supercell}=[0,0]$.

The RPA calculations in Fig.~\ref{fig2} (b) are much faster, but lose a little accuracy, compared to the self-consistent HFMF calculations in Fig.~\ref{fig2} (c). This is because all the interacting effects in our method only enter into matrix $A$, as shown in Eq.~\eqref{mutipole_exp} and Eq.~\eqref{mutipole_responce3}. The interacting wave-functions are not required and those time-consuming self-consistent HFMF calculations are avoided. Therefore, the phase transition points in Fig.~\ref{fig2} (b) ($U_{\text{rpa}}$) and Fig.~\ref{fig2} (c) ($U_{\text{sc}}$) cannot match exactly, as shown in Table \ref{tabs1}. In Fig.~\ref{fig2} (b), $U_{\text{rpa}}$ are identified by $\nu_{m}=0$, which are labeled by dashed lines. In Fig.~\ref{fig2} (c), $U_{\text{sc}}$ are identified at which the energy of the corresponding magnetic states become lower than NM state.

\begin{table}[t]
\newcommand{\tabincell}[2]{\begin{tabular}{@{}#1@{}}#2\end{tabular}}
\centering
\caption{Comparison between the results obtained with RPA calculations and self-consistent HFMF calculations. $U_{\text{rpa}}$ are the phase transition points identified by RPA calculations with $\nu_{m}=0$ in Fig.~\ref{fig2} (b). $U_{\text{sc}}$ are the phase transition points identified by self-consistent calculations in Fig.~\ref{fig2} (c), at which the energy of magnetic states become lower than NM state.}
\setlength{\tabcolsep}{2.0mm}{
\begin{tabular}{c|c|c|c|c|c|c|c}
\hline\hline
& Zigzag & Stripe & FM-Y & Oc-I & Oc-II & Oc-III & Oc-IV\\
\hline
$U_{\text{rpa}}$ (eV) & 0.40 & 0.50 & 0.79 & 3.10 & 3.10 & 3.33 & 3.30\\
\hline
$U_{\text{sc}}$ (eV) & 0.4 & 0.5 & 0.8 & 1.15 & 1.15 & 1.09 & 1.09\\
\hline\hline
\end{tabular}}
\label{tabs1}
\end{table}

\section{Analysis of $O^{33}_{21}$ and $O^{36}_{21}$ magnetic octupole moment}\label{appf}
\setcounter{figure}{0}
\setcounter{equation}{0}

According to Ref.~\cite{raab2005}, the magnetostatic field produced by the steady currents can be expanded in terms of the magnetic dipoles and quadrupoles as following (The repeated subscript denotes a summation),
\begin{equation}\label{mtpB}
\begin{aligned}
B_{i}(\mathbf{R})=& \frac{\mu_{0}}{4 \pi}\left[\frac{3 R_{i} R_{j}-R^{2} \delta_{i j}}{R^{5}} m_{j}+\frac{3}{2 R^{7}}\left\{5 R_{i} R_{j} R_{k}-R^{2}\left(R_{i} \delta_{j k}+R_{j} \delta_{k i}+R_{k} \delta_{i j}\right)\right\} m_{j k}+\cdots\right],
\end{aligned}
\end{equation}
where, $\mathbf{R}$ labels the coordinate of the field point with $i$, $j$ and $k$ denoting its $x$, $y$ or $z$ component, $\mathbf{m}$ labels the magnetic dipole, $m_{jk}$ labels the magnetic quadrupole. They are given by
\begin{equation}
\mathbf{m}=\sum_{\alpha=1}^{N} \frac{q^{\alpha}}{2 m^{\alpha}} \mathbf{L}^{\alpha}, \quad
m_{j k}=\sum_{\alpha=1}^{N} \frac{2 q^{\alpha}}{3 m^{\alpha}} L_{j}^{\alpha} r_{k}^{\alpha},
\end{equation}
where, the summations represent the moment contributions from different charge $q^{\alpha}$ with angular momentum $L^{\alpha}$ at position $\bm{r}^{\alpha}$. Similarly, we can generalize Eq.~\eqref{mtpB} up to third order,
\begin{equation}\label{mtpB3}
\begin{aligned}
B_{i}^{(3)}(\mathbf{R})=&\frac{\mu_{0}}{4 \pi} \frac{m_{j k l}}{2 R^{9}}\Big\{35 R_{i} R_{j} R_{k} R_{l}-5 R^{2}\left(R_{i} R_{j} \delta_{k l}+R_{j} R_{k} \delta_{i l}+R_{j} R_{l} \delta_{i k}+
R_{k} R_{l} \delta_{i j}+R_{i} R_{l} \delta_{k j}+R_{i} R_{k} \delta_{l j}\right)\\
&+R^{4}\left(\delta_{i j} \delta_{k l}+\delta_{k j} \delta_{i l}+\delta_{l j} \delta_{i k}\right)\Big\},
\end{aligned}
\end{equation}
and then get the magnetic octupole $m_{j k l}$ as
\begin{equation}\label{mtpcalssical}
m_{j k l}=\sum_{\alpha=1}^{N} \frac{3 q^{\alpha}}{4 m^{\alpha}} L_{j}^{\alpha} r_{k}^{\alpha} r_{l}^{\alpha}.
\end{equation}

\begin{figure}[t]
\centering
\includegraphics[width=0.4\textwidth,trim=0 0 0 0,clip]{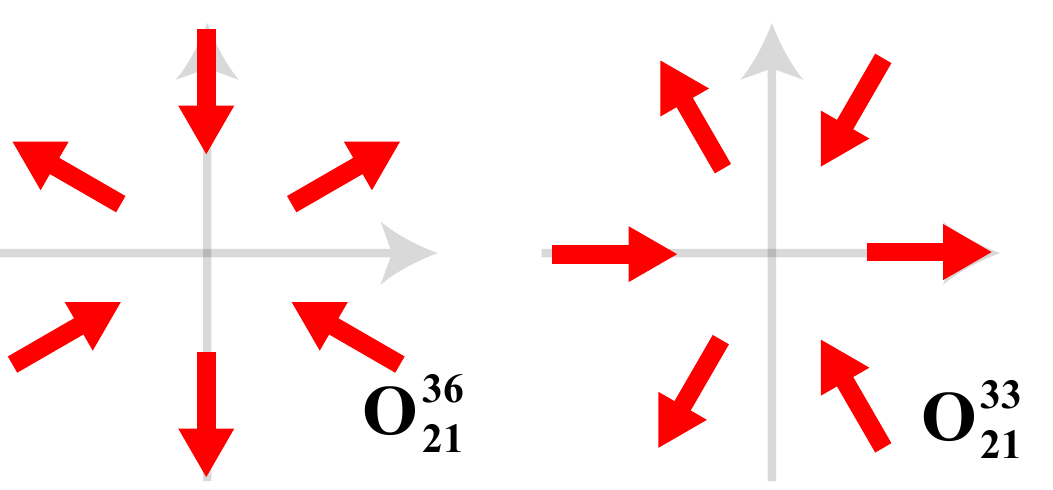}
\caption{(Color online). Distribution of magnetic field produced by $O_{21}^{36}$ and $O_{21}^{33}$ octupoles.
}
\label{figs2}
\end{figure}

The $O^{33}_{21}$ and $O^{36}_{21}$ octupoles defined in Ref.~\cite{yilin2017} are given by $\left(l_{x}^{2}-l_{y}^{2}\right) s_{x}-\left(l_{x} l_{y}+l_{y} l_{x}\right) s_{y}$ and $\left(l_{x}^{2}-l_{y}^{2}\right) s_{y}+\left(l_{x} l_{y}+l_{y} l_{x}\right) s_{x}$, respectively, where $\bm{l}$ and $\bm{s}$ label the orbital and spin angular momenta. According to Sec.~III of Ref.~\cite{fazekas1999}, we can understand these expressions by replacements of $l_j l_k+l_k l_j \leftrightarrow 2 r_j r_k$ and $s_i  \leftrightarrow L_i$ without changing its symmetry, namely $O_{21}^{33}\propto\left(r_{x}^{2}-r_{y}^{2}\right) L_{x}- 2 r_{x} r_{y} L_{y}$ and $O_{21}^{36}\propto\left(r_{x}^{2}-r_{y}^{2}\right) L_{y}+2 r_{x} r_{y} L_{x}$. Based on Eq.~\eqref{mtpB3} and Eq.~\eqref{mtpcalssical}, the magnetic field produced by these two octupoles can be written as
\begin{equation}
\mathbf{B}_{21}^{33}\propto\left\{\begin{array}{c}
B_{x}=7 R_{x}^{4}- 21 R_{x}^{2} R_{y}^{2} +3 R^{2} \left(R_{y}^{2}-R_{x}^{2}\right) \\
\quad \\
B_{y}= \left(7 R_{x}^{2}-21 R_{y}^{2}+6 R^{2}\right) R_{x} R_{y}\\
\quad \\
B_{z}=\left(7 R_{x}^{2}-21 R_{y}^{2}\right) R_{x} R_{z}
\end{array}\right. ,
\quad
\mathbf{B}_{21}^{36}\propto\left\{\begin{array}{c}
B_{x}=\left(-7 R_{y}^{2}+21 R_{x}^{2}-6 R^{2}\right) R_{x} R_{y} \\
\quad \\
B_{y}=-7 R_{y}^{4}+ 21 R_{x}^{2} R_{y}^{2} +3 R^{2} \left(R_{y}^{2}-R_{x}^{2}\right) \\
\quad \\
B_{z}=\left(-7 R_{y}^{2}+21 R_{x}^{2}\right) R_{y} R_{z}\quad
\end{array}\right. .
\end{equation}

\begin{figure}[htbp]
\centering
\includegraphics[width=0.6\textwidth,trim=0 0 0 0,clip]{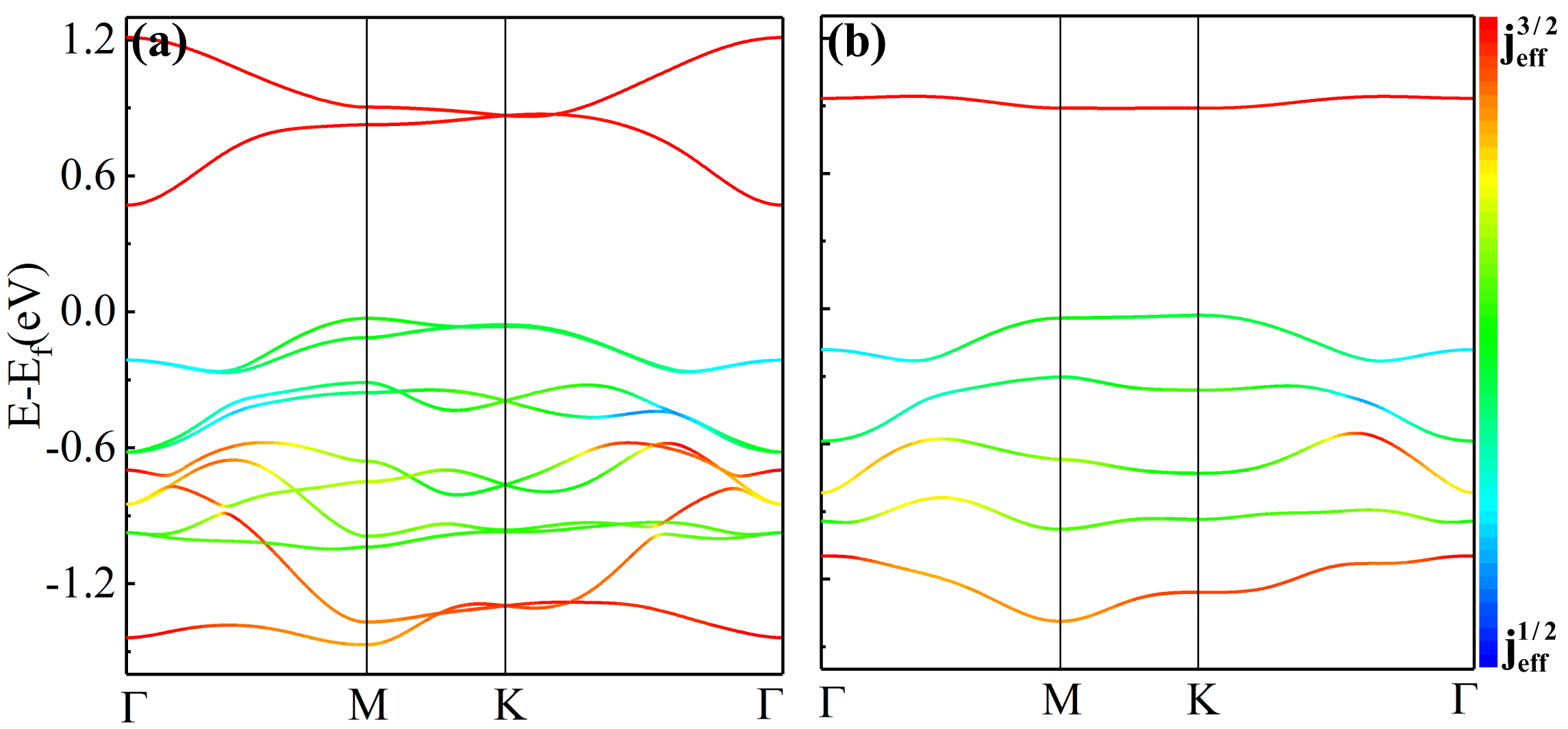}
\caption{(Color online). Band structures of $O_{21}^{36}$ states in $\alpha$-\ce{RuCl3}. $U=2.4$ eV, $J=0.4$ eV and $\lambda=96$ meV are used. (a) FM-$O_{21}^{36}$ state. (b) AFM-$O_{21}^{36}$ state.
}
\label{figs3}
\end{figure}

We can see that the field generated by $O^{33}_{21}$ is the counterpart of that generated by $O_{21}^{36}$ through an operation of $R_{x} \leftrightarrow -R_{y}$ and $B_{x} \leftrightarrow -B_{y}$. In Fig.~\ref{figs2}, we plot the field distributions in $xy$ plane. It shows that $O_{21}^{36}$ octupole satisfies $\textrm{C}_\textrm{3z}$ and $\textrm{C}_\textrm{2y}$ symmetries that belong to the point group of non-interacting Hamiltonian $H_0$ in monolayer $\alpha$-\ce{RuCl3}. Thus, it is possible for $\alpha$-\ce{RuCl3} to enter two kinds of pure meta-stable octupolar states of $O_{21}^{36}$. One is FM-$O_{21}^{36}$ state, in which the $H^{\mathrm{MF}}_{\mathrm{int}}$ with the ordered $O_{21}^{36}$ octupole only breaks TRS of $H_0$. Another is AFM-$O_{21}^{36}$ state with broken TRS and P symmetries (preserve PT). So in these two states, the total Hamiltonian $H_0+H^{\text{int}}_{\text{MF}}$ satisfies both $\textrm{C}_\textrm{3z}$ and $\textrm{C}_\textrm{2y}$, which can forbid dipole moments. The band structures of FM- and AFM-$O_{21}^{36}$ states are shown in Fig.~\ref{figs3}. Due to the PT symmetry, the bands of AFM-$O_{21}^{36}$ states remain doubly degenerate at each momentum $k$.

As a counterpart of $O_{21}^{36}$, the $O_{21}^{33}$ octupole satisfies $\textrm{C}_\textrm{3z}$ and $\textrm{C}_\textrm{2x}$ symmetries. However, $\textrm{C}_\textrm{2x}$ does not exist in $\alpha$-\ce{RuCl3}. Thus, when $O_{21}^{33}$ octupole appears, the system only has $\textrm{C}_\textrm{3z}$ and P (PT) symmetries, which are the same as the FM-Z (AFM-Z) ordered state. Therefore, $O_{21}^{33}$ octupole usually coexists with the magnetic dipoles that align along the $z$-direction. When $J\ge0.22$ eV, the $O^{33}_{21}$ state can not be stabilized anymore because the FM magnetic order is more favored by $J$. For clarity, we list the remaining symmetries with respected to $H_0$ in Table \ref{tabs2} for the magnetic ordered states identified in Fig.~\ref{fig2} (b) of the main text.

Now we discuss the detail about the experimental detection for $O_{21}^{36}$ state. The general formula of free energy and magnetization in a magnetic ordered state under an external magnetic field are given by~\cite{tian2017}
\begin{equation}
\left\{\begin{array}{c}
F=-\mathrm{M}^{\mathrm{d}} \cdot \mathrm{H}-\chi_{i j}^{\mathrm{P}} H_{i} H_{j}-Q_{i j} H_{i} H_{j}-\omega_{i j k} H_{i} H_{j} H_{k} \\
\quad \\
M_{i}=-\partial F / \partial H_{i}=M_{i}^{\mathrm{d}}+\chi_{i j}^{\mathrm{p}} H_{j}+Q_{i j} H_{j}+\omega_{i j k} H_{j} H_{k}
\end{array}\right.,
\end{equation}
where, $M_{i}^{\mathrm{d}}$ is the magnetic dipolar term, $Q_{ij}$ is the magnetic quadrupolar term, $\omega_{ijk}$ is the magnetic octupolar term and ${\chi}^{p}_{ij}$ is the paramagnetic term. In the $O_{21}^{36}$ state, the free energy and magnetization contributed by $O_{21}^{36}$ octupole take the following form enforced by its $\textrm{C}_\textrm{3z}$ and $\textrm{C}_\textrm{2y}$ symmetries,
\begin{equation}
\left\{\begin{array}{c}
F_{21}^{36}=\omega_{21}^{36}\left\{\left(H_{x}^{2}-H_{y}^{2}\right) H_{y}+2 H_{x}^{2} H_{y}\right\} \\
\quad \\
M_{x}=6 w_{21}^{36} H_{x} H_{y},\quad
M_{y}=3 w_{21}^{36}\left(H_{x}^{2}-H_{y}^{2}\right),\quad
M_{z}=0
\end{array}\right. ,
\end{equation}
where $\omega^{36}_{21}$ is the octupolar susceptibility. Thus, an orthogonal magnetization oscillation of $M_y\propto H^2\cos^2\theta$ would be expected if a rotating magnetic field $H$ is applied in the $xz$ plane with $\theta$ respect to $x$-axis.

Note that $Q_{ij}$ is forbidden by local inversion symmetry preserved by all multipoles composed of $l$ and $s$~\cite{yilin2017}, and there is also no $M_{i}^{\mathrm{d}}$ in $O_{21}^{36}$ state as mentioned above. The paramagnetic moments contributed by ${\chi}^{p}_{ij}$ are parallel to $H$. Once the orthogonal magnetization $M_y$ is observed, it is a direct evidence of the existence of $O_{21}^{36}$ octupole.

\begin{table}[t]
\newcommand{\tabincell}[2]{\begin{tabular}{@{}#1@{}}#2\end{tabular}}
\centering
\caption{Symmetry breaking with respect to non-interacting Hamiltonian $H_0$ of monolayer $\alpha$-\ce{RuCl3} for states identified in Fig.~\ref{fig2} (b) and Table \ref{tab1} in our main text.}
\setlength{\tabcolsep}{2mm}{
\begin{tabular}{c|c|c|c|c|c|c|c}
\hline\hline
\diagbox{Symmetry}{${O}^{\text{eig}}_{m}$} & Zigzag & Stripe & FM-Y & FM-$O_{21}^{33}$ & FM-$O_{21}^{36}$ & AFM-$O_{21}^{33}$ & AFM-$O_{21}^{36}$\\
\hline
TRS & \XSolidBold & \XSolidBold & \XSolidBold & \XSolidBold & \XSolidBold & \XSolidBold & \XSolidBold\\
\hline
E & \CheckmarkBold & \CheckmarkBold & \CheckmarkBold & \CheckmarkBold & \CheckmarkBold & \CheckmarkBold & \CheckmarkBold\\
\hline
P & \XSolidBold & \CheckmarkBold & \CheckmarkBold & \CheckmarkBold & \CheckmarkBold & \XSolidBold & \XSolidBold\\
\hline
$\textrm{C}_\textrm{3z}$ & \XSolidBold & \XSolidBold & \XSolidBold & \CheckmarkBold & \CheckmarkBold & \CheckmarkBold & \CheckmarkBold\\
\hline
$\textrm{C}_\textrm{2y}$ & \XSolidBold & \XSolidBold & \CheckmarkBold & \XSolidBold & \CheckmarkBold & \XSolidBold & \CheckmarkBold\\
\hline\hline
\end{tabular}}
\label{tabs2}
\end{table}

\section{Negative SOC and octupolar ground state in monolayer $\alpha$-$\mathrm{RuI_3}$}\label{appg}
\setcounter{figure}{0}
\setcounter{equation}{0}

\begin{figure}[t]
\centering
\includegraphics[width=0.9\textwidth,trim=0 0 0 0,clip]{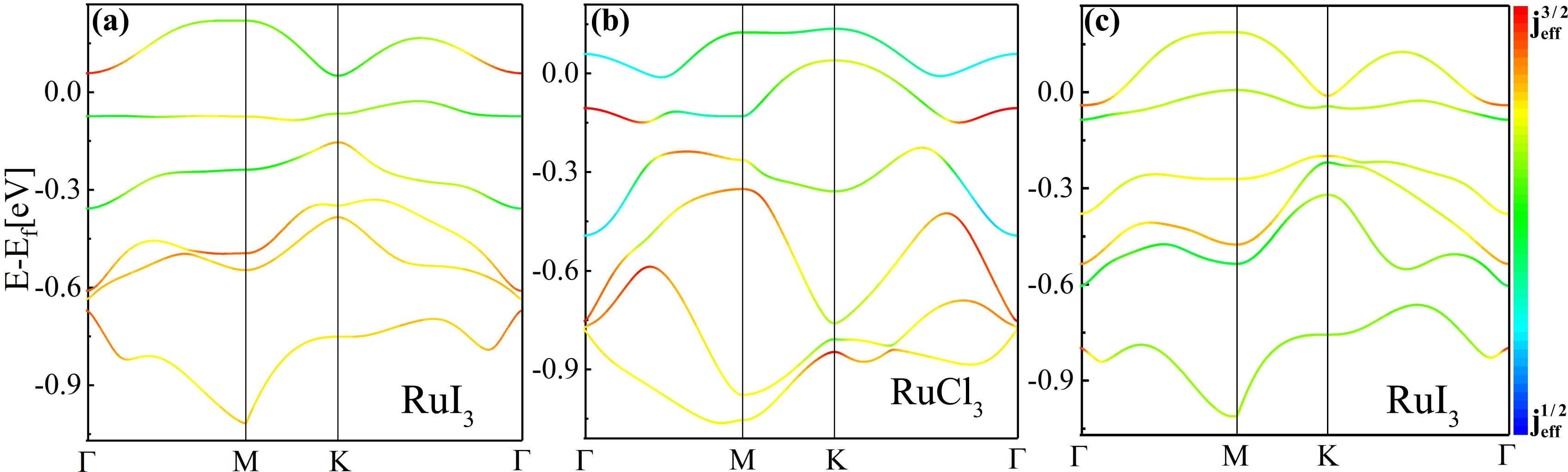}
\caption{(Color online). Non-magnetic band structures obtained by DFT calculations directly with SOC for monolayer $\alpha$-\ce{RuI3} (a) and $\alpha$-\ce{RuCl3} (b). (c) Non-magnetic band structures for monolayer $\alpha$-\ce{RuI3}, where the non-SOC part $H_0$ is constructed by DFT calculations combined with the maximally localized Wannier functions method. A SOC term of $\lambda\bm{l}\cdot\bm{s}$ is added to $H_0$ with $\lambda=-100$ meV.
}
\label{figs4}
\end{figure}

According to $-\lambda \bm{l}_{t_{2g}} \cdot \bm{s} = \lambda (-\bm{l}_{t_{2g}})\cdot \bm{s}=\lambda \bm{l}_{p}\cdot \bm{s}$, a negative SOC can be obtained by replacing $\textrm{Cl}^\textrm{-}$ with $\textrm{I}^\textrm{-}$, through which the more extended $p$ orbitals of $\textrm{I}^\textrm{-}$ with stronger negative SOC can be much more mixed with $t_{2g}$ orbitals of $\textrm{Ru}^\textrm{3+}$~\cite{sheng2014,nie2017}. Based on the optimized structure, we calculate the non-interacting band structures of monolayer $\alpha$-\ce{RuI3} by DFT calculations with SOC included, as shown in Fig.~\ref{figs4} (a), in which the color-bar shows orbital projection of $j_\textrm{eff}=\frac{3}{2}, \frac{1}{2}$. For comparison, we also give the band structures of $\alpha$-\ce{RuCl3} in Fig.~\ref{figs4} (b). It obviously shows that the bands of $\alpha$-\ce{RuI3} around the Fermi level have more green and yellow characters ($j_\textrm{eff}=\frac{3}{2}$) compared with that of $\alpha$-\ce{RuCl3}, which implies that the mixing of $p$ orbitals has significantly influenced the $\lambda$. We then calculate the non-interacting band structures of $\alpha$-\ce{RuI3} by non-SOC DFT calculations, as shown in Fig.~\ref{figs1} (d)-(e). Fig.~\ref{figs1} (f) is the projected density of states, which confirms that more $p$ components are mixed with $t_{2g}$ orbitals compared to that in $\alpha$-\ce{RuCl3} (Fig.~\ref{figs1} (c)). The non-SOC part $H_0$ is constructed by the maximally localized Wannier functions method and a SOC term of $\lambda\bm{l}\cdot\bm{s}$ is added to $H_0$. The resulting band structures are plotted in Fig.~\ref{figs4} (c) with $\lambda=-100$ meV. We can see that the band shape and $j_\textrm{eff}=\frac{3}{2}$ character around the Fermi level are qualitatively in agreement with that from DFT calculations directly (Fig.~\ref{figs4} (a)), which strongly suggests that the negative SOC of $t_{2g}$ bands can be achieved in $\alpha$-\ce{RuI3}.

\begin{figure}[t]
\centering
\includegraphics[width=0.8\textwidth,trim=0 0 0 0,clip]{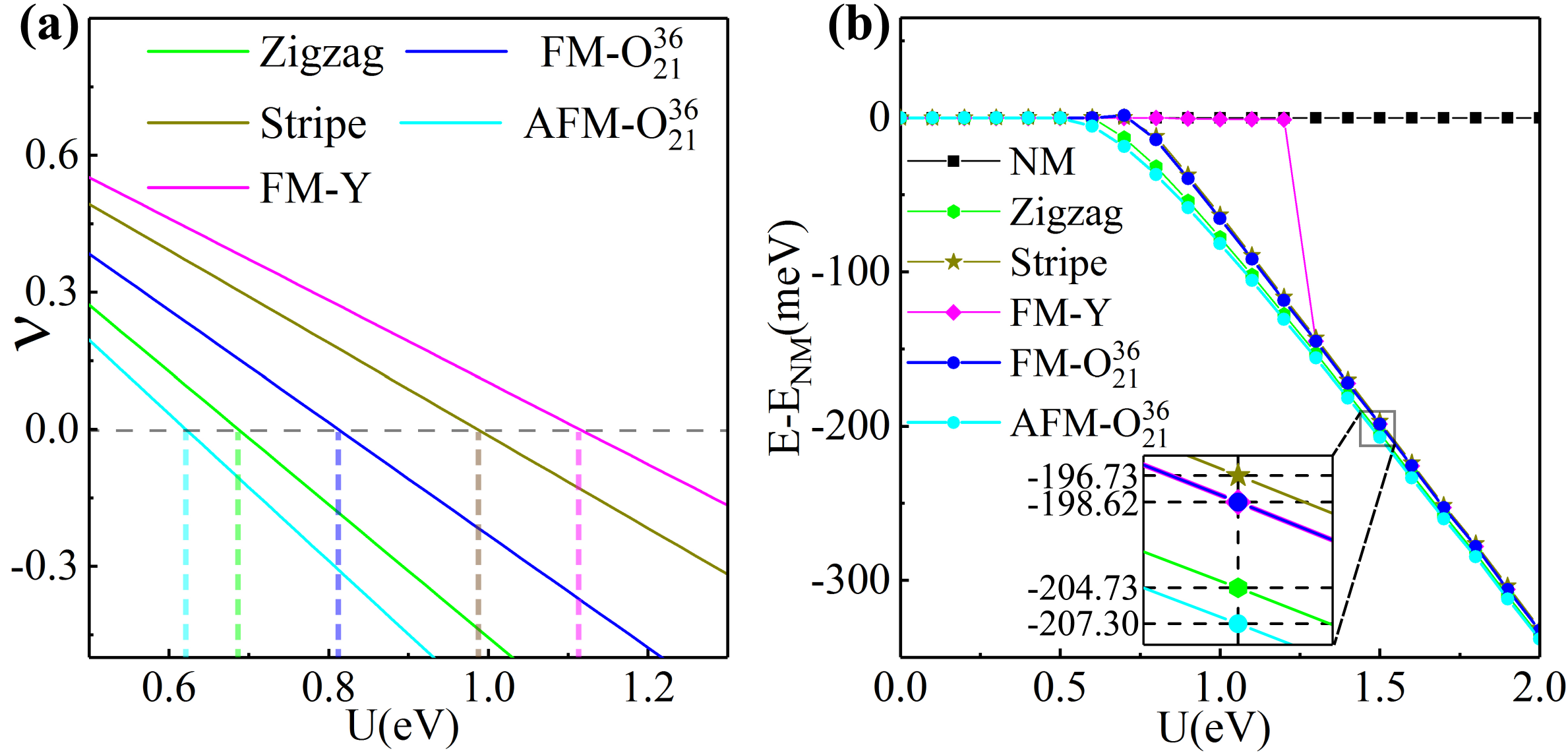}
\caption{(Color online). Magnetic states calculated in monolayer $\alpha$-\ce{RuI3}. (a) The eigenvalues $\nu_{m}$ of $(I-{\chi}^{0} A)$ that approach to zero as a function of $U$. (b) Total energy of magnetic states relative to the NM state as a function of $U$. The inset shows that AFM-$O^{36}_{21}$ state has the lowest energy. $J=0$ eV and $\lambda=-100$ meV are used in (a)-(b).}
\label{figs5}
\end{figure}

We further calculate the magnetic states of $\alpha$-\ce{RuI3} based on $\lambda=-100$ meV using the same method as $\alpha$-\ce{RuCl3}. Fig.~\ref{figs5} (a) shows the magnetic states identified by RPA calculations. Fig.~\ref{figs5} (b) shows their energy relative to NM state obtained by self-consistent unrestricted HFMF calculations. We can see that the AFM-$O^{36}_{21}$ state has the lowest energy, which is about 3 meV lower than the Zigzag dipolar order. Therefore, we think it is possible to realize such octupolar ground state in monolayer $\alpha$-\ce{RuI3}.

\twocolumngrid

\end{document}